# Measurement of the Hall effect at nanoscale with three probes

G. X. Chen, R. X. Cao, A. Zholud, and S. Urazhdin[*]

*Department of Physics, Emory University, Atlanta, Georgia 30322, USA*

*Corresponding author: surazhd@emory.edu

**Abstract**

The Hall effect and its varieties such as quantum, anomalous, and spin Hall effects, provide indispensable tools for the characterization of electronic and magnetic properties of materials, metrology, and spintronics. The conventional four-probe Hall configuration is generally not amenable to measurements at nanoscale, due to current shunting by the Hall electrodes. We demonstrate that Hall measurements on the nanoscale can be facilitated by the three-probe Hall configuration that avoids the shunting problem. We illustrate the efficiency of the proposed approach with anomalous Hall effect-based measurements of individual activation events during domain wall motion in magnetic films with perpendicular anisotropy.

## I. Introduction

Since the discovery of the Hall effect in 1879, it has found many scientific and technological applications in electronic and magnetic characterization of materials, sensing, and metrology.[1-3] The observation of the quantized Hall effect in two-dimensional materials has transformed our understanding of the phases of matter, ushering in a new paradigm of topological phase classification.[4, 5] Hall effects caused by the spin-orbit interaction instead of the Lorentz force – the spin Hall effect and the anomalous Hall effects – have become indispensable for the spintronics and magnetism research.[6-9]

Measurements of the Hall effect generally involve two pairs of electrodes. One pair is utilized to inject electrical current, while the second pair, transverse to the first one, is utilized for the Hall voltage measurement. Nanopatterned Hall bars are generally utilized in micrometer-scale measurements.[10-13] This approach cannot be extended to structures with dimensions below a few hundred nanometers, because the electronic properties of the Hall bar edges are modified by the patterning, and their effects at nanoscale can overwhelm the bulk contribution. This problem would be avoided in a Hall measurement utilizing four electrodes deposited on an extended film, but the current flow for such a pattern would be disrupted by shunting through the Hall electrodes. To illustrate this point, in Fig.1(a) we show the distribution of current calculated for such a Hall measurement geometry using the COMSOL simulation software. The electronic properties and the thickness of the film in this calculation were chosen to match those of the material used in the measurements discussed below. The endpoints of the 80 nm-thick Au Hall voltage and current electrodes are separated by 150 nm, and their radius of curvature is 50 nm, typical for nanopatterns defined by electron beam lithography. The x-axis is directed along the average current flow, and the y-axis is in the film plane, normal to the flow. As Fig.1(a) shows, the current spreads towards the voltage electrodes, and its density becomes noticeably reduced in the region between these electrodes, due to shunting. These qualitative observations are confirmed by the analysis of the current distribution through the center of the structure along the y-axis, Fig.1(b). The current distribution in the film is shown with a solid curve, and that in the electrodes – with a dotted curve. The current flowing in film between the voltage electrodes, which determines the Hall voltage, constitutes only about 1/4 of the total current, while the remaining current is shunted through the leads. We conclude that shunting can significantly reduce the Hall signal, by the amount that depends on the conductivity of the studied material and the electrodes, as well as the details of the measurement geometry.

## II. Three-probe Hall measurement approach

We propose and experimentally demonstrate that the three-probe Hall technique[14, 15] can be adopted for the Hall measurements at nanoscale. It does not suffer from the drawbacks of the four-probe Hall measurement discussed above, and can be arbitrarily downscaled without compromising the measurement. In the proposed approach, three electrodes are positioned symmetrically around the probed area of the film, as shown in Fig.1(c). The same ac or dc current $I$ is sourced by two of the electrodes, and current $2I$ is drained from the third electrode. This can be accomplished with two identical current sources, or with a single voltage source and two resistors with values significantly larger than the sheet resistance of the film, as shown in Fig.1(c).

In the absence of the Hall effect, the symmetry of the system dictates that the voltage $V_1$ between the first and the third electrode is equal to the voltage $V_2$ between the second and the third electrode. However, in the presence of the Hall effect, these voltages are generally not equal. Therefore, $V_{12}=V_1-V_2$, the voltage between the first and the second electrodes, represents a quantitative measure of the Hall effect in the system. Measurements described below confirm a direct relation between $V_{12}$ obtained in the three-probe geometry, and the Hall voltage measured in the standard four-probe geometry.

When thick high-conductivity Au or Cu electrodes are used, their sheet resistance is typically significantly smaller than that of the studied materials. In the 4-probe Hall geometry, higher electrode conductivity leads to stronger shunting. In contrast, simulations show that in the proposed 3-probe geometry, high conductivity of electrodes helps confine the current flow mostly to the area defined by the tips of the three electrodes, as shown in the map Fig.1(d) and with a dashed curve in the crossection Fig.1(b). Thus, the proposed configuration enables Hall probing at nanoscale, without compromising the Hall signal by shunting.

The tradeoffs between the three-probe Hall measurement and the standard four-probe measurement can be understood by analogy with the two-probe vs four-probe resistance measurements. Specifically, only two-probe resistance measurements are feasible in structures with dimensions below a few hundred nanometers, because of the detrimental effects of shunting in the 4-probe approach, similar to those discussed above. In contrast to the four-probe approach, two-probe resistance measurements are affected by the contact resistances and the resistances of electrical leads. In three-probe Hall measurement, the voltage $V_{12}$ is not affected by either the contact or the electrical lead resistance, as long as these resistances are the same for both electrodes 1 and 2. However, an asymmetry between these contributions, and/or the geometric asymmetry between the electrodes is expected to result in an additional contribution to the measured voltage. This contribution can be eliminated by averaging over the three possible equivalent choices of the third electrode.

**III. Samples and experimental details**

We experimentally verified the viability of the proposed nanoscale three-probe Hall configuration using magnetic films with perpendicular magnetic anisotropy (PMA). At modest magnetic fields, the Hall voltage for such films is dominated by the anomalous Hall effect (AHE), which results from the spin-orbit coupling instead of the Lorentz force, and reflects the configuration of the magnetization in the system.

AHE has been instrumental in many recent studies of spin-orbit interaction in magnetic films and at their interfaces. The related spin Hall effect, observed in nonmagnetic materials, has emerged as an efficient source of spin currents that does not rely on the spin-filtering properties of ferromagnets.[16] In this context, the proposed three-probe measurement can become an indispensable tool for the studies of nanoscale phenomena such as the recently discovered nanoscale magnetic textures, the skyrmions.[17]

The AHE voltage is proportional to the component of magnetization perpendicular to the film plane,[18,19] and can be described by the empirical formula: $V_H = \frac{I}{t}(R_{OH}B_z + \mu_0 R_{AHE} M_z)$,[20] where $M_z$ is the component of the magnetization normal to the film plane, $R_{AHE}$ is the anomalous Hall coefficient determined by the band structure and electron scattering, $I$ is the current, and $t$ is the thickness of magnetic film. The first term in parentheses represents the ordinary Hall contribution described by the Hall coefficient $R_{OH}$. At modest fields, it is usually negligible compared to the AHE.

The studied samples were based on the Ta(4)Pt(5)[Co(0.8)Pt(0.4)]$_{\times 6}$Pt(1.6) miultilayer films deposited by high-vacuum magnetron sputtering at room temperature. All thicknesses are given in nanometers. The Ta(4)Pt(5) bilayer served as a buffer to promote the PMA of the CoPt magnetic multilayer, and the Pt(1.6) capping layer protected the magnetic layer from oxidation. To enhance the PMA of the CoPt multilayer, the films were annealed at 220° C for 40 minutes, in magnetic field B=0.3 T oriented normal to the film surface.[21] The magnetoelectronic measurements were performed at room temperature, using an ac driving current of 0.045 mA rms and lock-in detection.

We performed Hall measurements using macroscopic films deposited on 6×6 mm$^2$ square chips, as well as nanofabricated samples. For the 4-probe Hall measurements, the macroscopic samples were contacted in the standard van der Pauw geometry. For the 3-probe measurements, three approximately 0.5 mm-sized indium dots were pressed on the surface of the sample at equal distances of about 5 mm from one another. To fabricate the nanoscale samples, the Pt/Co multilayer was patterned by e-beam lithography (EBL) into a 6-μm disk. Three sharp Au(80) electrodes, arranged symmetrically around the disk, were then patterned by EBL and thermal evaporation. The distance between the tips of the electrodes was 200 nm, as verified by the scanning electron microscopy, inset in Fig.2(a).

**IV. Results**

Magnetoelectronic hysteresis loops of the macroscopic CoPt film, obtained in both 4-probe and 3-probe Hall geometries, are shown in Fig.2(a). These data demonstrate that the two measurements are equivalent for the macroscopic samples. Solid curves in Fig.2(b) show several hysteresis loops for the nanoscale sample, acquired sequentially using the 3-probe AHE measurement, together with the hysteresis loop for the macroscopic sample shown by dashed curves.

Several differences between the hysteresis loops of the nanoscale and the macroscopic samples can be identified. First, the curves for the macroscopic sample are smooth, except for the initial jump from the saturated state typical for the "bowtie" hysteresis loop in high-quality PMA films. Meanwhile, the hysteresis loops for the nanoscale sample are dominated by several abrupt jumps. Second, the hysteresis loop of the macroscopic sample is reproducible in the repeated measurements (not shown), while for the nanoscale sample, the height of the individual jumps, and the field at which they occur, are different in each loop. Finally, aside from the jumps, the hysteresis loops of the nanoscale sample are somewhat broader. The latter may be associated with the additional annealing unavoidable in the e-beam lithography procedure utilized in nanoscale sample fabrication.

The jumps observed for the nanoscale sample can be attributed to the Barkhausen effect[22], i.e. individual activation events associated with the magnetization reversal process that occurs by the nucleation of reversed domains, and their subsequent expansion via thermally activated motion of domain walls (DWs).[23,24] The fact that only a few jumps are observed in the hysteresis loop suggests that only a few strong pinning cites are present in the 200-nm circular sample area that dominates the measured signal.

The variations of the positions and heights of the Hall signal jumps in the repeated measurements for the nanoscale sample Fig.3(b) are consistent with the random probabilistic nature of the DW propagation. For the macroscopic film, the random variations become averaged out, resulting in a smooth and reproducible hysteresis loop. Since the mechanism underlying these dynamics is random thermal activation, such jumps/evolution should occur even at a fixed field, as the DWs relax towards their stable configuration. To verify the existence of this magnetic aftereffect, we have performed additional measurements of the time-dependent 3-probe Hall signal for the nanoscale samples. First, a saturating field of $H_z$=1.5 kOe perpendicular to the film was applied for 20 seconds. The field was then abruptly changed to -160 Oe, and the time-dependent Hall voltage was recorded over a period of 1200 s. The final field was chosen to be slightly above that corresponding to the initial jump in the hysteresis loop associated with the onset of nucleation and propagation of DWs. The measurement was repeated 3 times for both the macroscopic and the nanoscale samples, as shown in Figs. 2(c),(d).

For the macroscopic film [Fig.3(c)], the Hall signal exhibits a smooth time dependence, and is reproducible, consistent with averaging over many individual DW activation events.[24,25] For the nanoscale sample [Fig.3(d)], the curves exhibit a few abrupt jumps with plateaus in between, consistent with the abrupt DW activation among a few pinned configurations. The height of the jumps and the time at which they occur, as well as the Hall voltage in the long-time limit, all vary among the sequentially acquired curves, consistent with the random probabilistic nature of both the DW motion and of the metastable magnetic configuration at long times. These data also confirm that the proposed 3-probe Hall measurement is exclusively sensitive to the magnetic configuration of the nanoscale region defined by the electrodes – otherwise, the data would also exhibit many smaller steps and/or gradual variation due to the magnetization evolution in the surrounding film.

**V. Summary**

We have proposed and experimentally verified the viability of the 3-probe Hall measurement at nanoscale, which overcomes the problem of current shunting in the conventional four-probe Hall geometry. The proposed approach enables Hall-effect characterization of local electronic and magnetic properties in nanometer-scale regions of unpatterned thin films, whose scale is limited only by the nanopatterning technique. The proposed configuration was experimentally verified by the characterization of magnetization reversal in Co/Pt multilayer films with perpendicular magnetic anisotropy. For a 200 nm region probed by the proposed technique, the hysteresis loop and the magnetic after-effect measurements at a constant field show abrupt jumps due to the random individual domain wall activation events. In contrast, similar measurements for macroscopic samples exhibit smooth curves consistent with the averaging over a large number of activation events. The proposed approach can provide a new tool for the studies of the nanoscale phenomena, such as current-induced static and dynamical local textures in magnetic films.[26,27]

## ACKNOWLEDGMENTS

We acknowledge support from the National Science Foundation grant No. DMR-1504449.

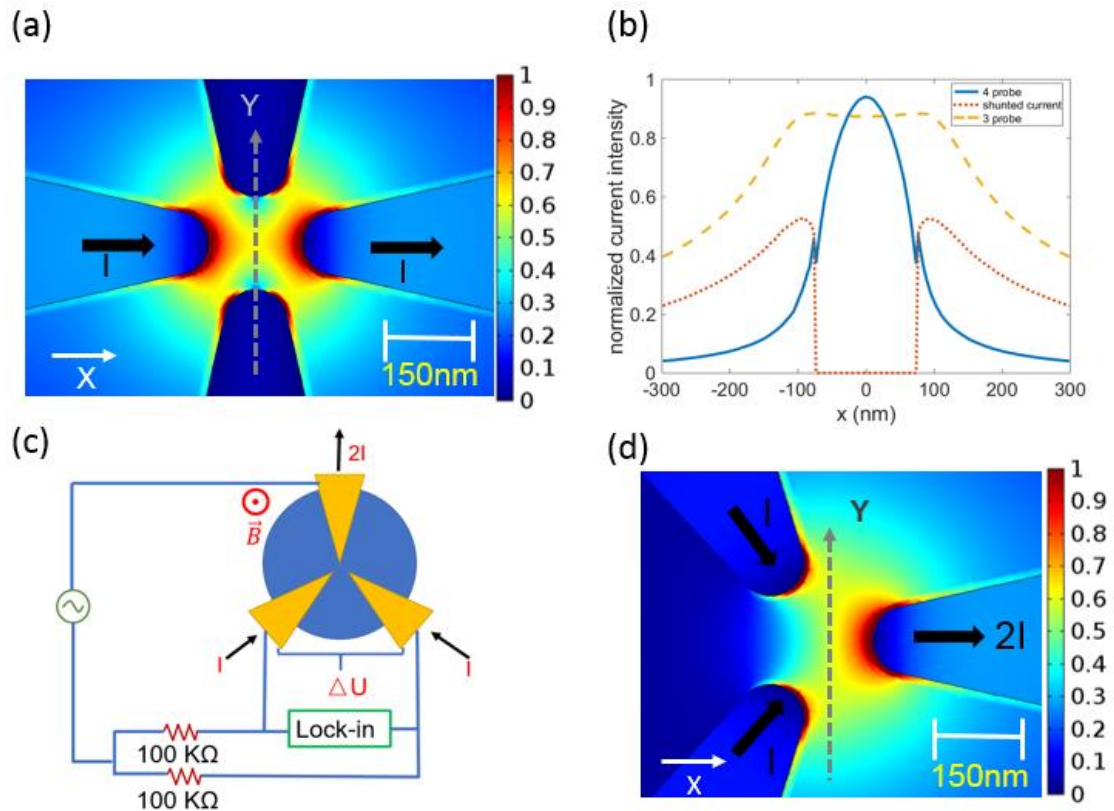

Fig.1. (a) Calculated distribution of current in a thin film with four Hall electrodes on top. The radius of curvature of the electrodes is 50 nm, and the tip-to-tip separation between the voltage and current electrodes is 150 nm. (b) Current distribution in the y-crossection through the center of the structure, for the film (solid curve) and the electrodes (dotted curve) in the 4-probe geometry, and for the 3-probe Hall geometry (dashed curve). (c) Schematic of the 3-probe Hall geometry and measurement setup. (d) Calculated current distribution in the film for the 3-probe Hall structure. The tip-to-tip separation between the electrodes is 150 nm. The calculations were performed with the COMSOL Multiphysics software, using $1.4 \times 10^7 S/m$ and $6.0 \times 10^7 S/m$ for the conductivities of the film and the electrodes, and 16 nm and 80 nm for the respective thicknesses.

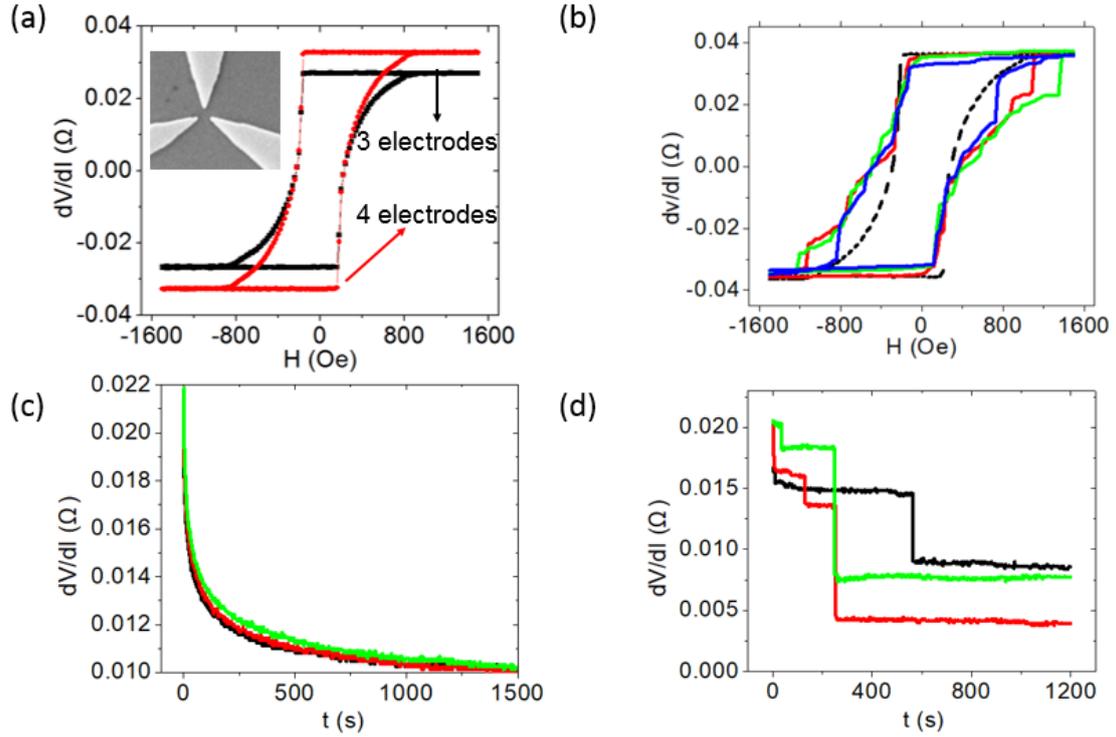

Fig. 2. (a) Hall hysteresis loops for a 6×6 mm$^2$ Ta(4)Pt(5)[Co(0.8)Pt(0.4)]$_{×6}$Pt(1.6) multilayer film, measured in the 3-probe (solid curve) and 4-probe (dashed curve) Hall geometries. The curves were vertically shifted to compensate for the resistive contribution caused by the electrode asymmetry. Inset: SEM micrograph of the nanoscale sample used for the 3-probe measurements shown in panels (b) and (d). The distance between the tips of electrodes is 200 nm. (b) Hysteresis loops of the nanoscale sample in repeated measurements (solid curves) and of the mm-scale extended film (dashed curve) with the same CoPt multilayer structure as in (a), in the three-probe geometry. To facilitate direct comparison, the hysteresis loop of the extended film is scaled with a factor of 0.8, and the curves are vertically shifted to compensate for the resistive contribution. (c), (d) Three sequentially acquired time dependencies of the 3-probe Hall signal vs time for the extended film (c) and for the nanoscale sample (d), acquired at H=-220 Oe and -160 Oe, respectively. The curves are vertically shifted by the same amount as in panels (a) and (b).